\documentclass[%
 reprint,
 nofootinbib,
 amsmath,amssymb,
 aps,
]{revtex4-2}
\usepackage[usenames,dvipsnames,table,xcdraw]{xcolor}
\usepackage[colorlinks,urlcolor=blue,citecolor=blue,linkcolor=blue, pdfborder={0 0 0}]{hyperref}
\usepackage[nameinlink,capitalise]{cleveref}
\usepackage{xspace}
\usepackage{booktabs}
\usepackage{xfp}
\usepackage{graphicx}% Include figure files
\usepackage{dcolumn}% Align table columns on decimal point
\usepackage{bm}% bold math
\begin{document}

% \preprint{APS/zQED}

\title{Amplitude modulation in binary gravitational lensing of gravitational waves}

\author{Yi Qiu$^{1,2,3}$}
\thanks{These authors contributed equally to this paper.}
\author{Ke Wang$^{4,5}$}
\thanks{These authors contributed equally to this paper.}
\author{Jian-hua He$^{1,2}$}
\thanks{Corresponding author: \href{mailto:hejianhua@nju.edu.cn}{hejianhua@nju.edu.cn}}
\affiliation{$^1$School of Astronomy and Space Science, Nanjing University, Nanjing 210093, P. R. China}
\affiliation{$^2$Key Laboratory of Modern Astronomy and Astrophysics (Nanjing University), Ministry of
Education, Nanjing 210023, P. R. China}
\affiliation{$^3$Department of Physics, The Pennsylvania State University, University Park PA 16802, USA}
\affiliation{$^4$Lanzhou Center for Theoretical Physics, Key Laboratory of Theoretical Physics of Gansu Province,\\
School of Physical Science and Technology, \\
Lanzhou University, Lanzhou 730000, P. R. China}
\affiliation{$^5$Institute of Theoretical Physics \& Research Center of Gravitation,\\ 
Lanzhou University, Lanzhou 730000, P. R. China}

\date{\today}

\begin{abstract}
We investigate the detectability of gravitational waves (GWs) lensed by a system that consists of binary black holes as lenses using time-domain numerical simulations. The gravitational lensing potential of this system is no longer static but evolves with time. When GWs from the source pass through the binary lens, their amplitudes can be modulated, which is similar to the phenomenon of amplitude modulation (AM) in radio communication. We find that even the frequency of the binary lens itself is too low to be detected by the LISA detection band, the sidebands in the spectrum of the lensed GWs due to AM can still be within the sensitive range of the detection band. Moreover, we also calculate the relative differences of SNR ({\it mismatch}) between the lensed and unlensed GWs.  We find that the {\it mismatch} can be as significant as 9.18\%. Since {\it mismatch} does not depend on the amplitude of wavefrom, the differences between the binary lensed and unlensed waveforms are substantial. This provides a robust way to identify the lensing event for the LISA project in the future.
\end{abstract}

\maketitle

{\bf Introduction} The discovery of gravitational waves (GWs) ushered us into a new area of astronomy. Similar to light, GWs can be lensed when they pass a massive object. Such a massive object forms a lensing system, which allows us to explore new physics which is beyond the scope of conventional astronomy, such as determining the location of merging black holes to subarcsecond precision~\cite{10.1093/mnras/staa2577} and detecting intermediate-mass or primordial black holes~\cite{PhysRevD.98.104029,PhysRevD.101.123512}. 

Most recently, a comprehensive analysis of lensing has been performed using the data from the first half of the third LIGO–Virgo observing run~\cite{LIGOScientific:2021izm}. The search includes strongly lensed events, multiple images, and microlensing effects. However, no compelling evidence of lensing has been found yet.

One reason for the null result is that the expected rate of lensing is low at the current detector sensitivities~\cite{LIGOScientific:2021izm}. However, another important reason is that the lensed templates used in the search are practically too close to the unlensed (GR) ones. The search focuses only on a single static lens. The waveforms of such a lens are based on the thin lens model~\cite{Schneider} as well as under the geometric optics approximation~\cite{Takahashi_2003}, in which the wavelength of GWs is assumed to be much smaller than the Schwarzschild radius $\lambda\ll 2M$ of the lens~\cite{Takahashi_2003,PhysRevD.98.104029}. As a result, the lensing effect changes only the amplitude and phase of the waveform. However, because of the degeneracy between the lensing magnification and the luminosity distance, lensing magnification alone can not be used to effectively identify a lensing event, unless additional information is available, such as the tidal effects in a binary neutron star system\cite{2020MNRAS.495.3740P}. Moreover, the lensing effect due to the phase shift is small as well. The relative differences of the signal-to-noise ratio (SNR) ({\it mismatch}) between the lensed and unlensed waveforms caused by phase shift are estimated less than $1\%$ for asymmetric binaries and less than $5\%$ for precessing and eccentric binaries with respect to the O3 sensitivity curves of the advanced LIGO and Virgo~\cite{PhysRevD.103.064047}.

When the wavelength of GWs is comparable to or much greater than the Schwarzschild radius $\lambda\gg 2M$ of the lens, wave effects become significant~\cite{He:2021hhl}. In this case, the geometric optics approximation and the thin lens model break down. GWs do not form caustics after the lens due to the wave effects at scales that are comparable to the wavelength. Instead, GWs form a strong beam along the optic axis~\cite{He:2021hhl}. However, despite such big differences, due to the degeneracy between the luminosity distances and lensing magnification, such wave effects, indeed, can not significantly enhance the detectability of the lensing events for a single static lens.  

In this {\it letter}, we investigate a new lensing system that hasn't been explored before. This new system consists of a binary lens, the potential of which is no longer static but evolves with time. When GWs from the source pass through the binary lens, the amplitude of GW signals can be modulated by the evolving potential, which is similar to the phenomenon of amplitude modulation (AM) in radio communication. By periodically changing the amplitude of GWs from source, the binary lens can leave detectable features in the spectrum of the lensed GWs. This may provide a new way to detect the lensing event in the future Laser Interferometer Space Antenna (LISA) project. Throughout this paper, we adopt the geometric unit $c=G=1$, in which $1\,{\rm Mpc}=1.02938\times 10^{14} {\rm Hz}^{-1}$ and $1 M_{\odot}=4.92535\times 10^{-6} {\rm Hz}^{-1}\,.$

\begin{figure}[]
\begin{center}
\includegraphics[width= 9cm]{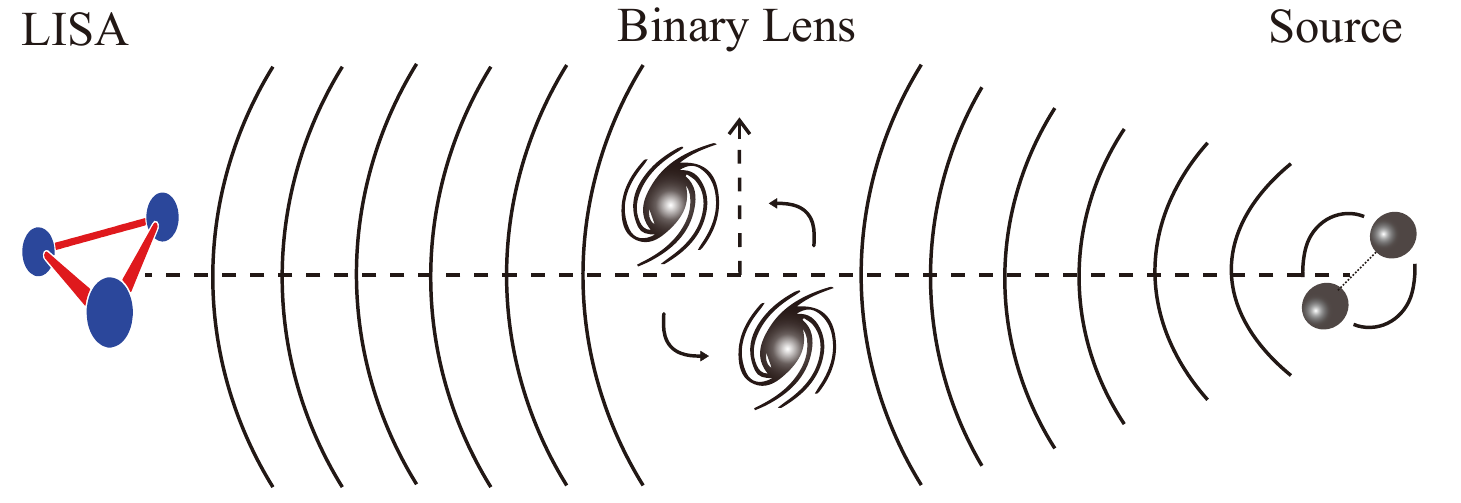}
\end{center}
\caption{Diagram of the binary lensing system. The signals of source GWs are generated by binary super-massive black holes with equal masses of $1\times 10^5 M_{\odot}$. The frequency of the source binary is in its inspiral phase at $5\times 10^{-4}$Hz. The generated GWs then pass through a binary lens with a total mass of $2.4\times 10^7 M_{\odot}$. The frequency of the inspiral binary lens is as low as $5\times 10^{-6}$Hz, which is outside the LISA detection band. Thus, the GWs generated by the binary lens itself can not be detected by LISA. However, through the amplitude modulation of the source GWs, the binary lens
can leave detectable features in the spectrum of the lensed GWs within the LISA detection band.
\label{fig:twobody}}
\end{figure}

{\bf The Model} We consider gravitational waves (GWs) propagating in a non-flat spacetime. In the weak field limit, the background field metric is given by
\begin{equation}
ds^2=-(1+2\psi)dt^2+(1-2\psi)d{\bf{r}} ^2\equiv g^{(\rm B)}_{\mu\nu}dx^{\mu}dx^{\nu}\,,
\end{equation}
where $\psi(t, {\bf{r}})\ll1$ is a time-dependent potential well. We consider a linear perturbation $h_{\mu\nu}$ on the background metric tensor $g^{(\rm B)}_{\mu\nu}$
\begin{equation}
g_{\mu\nu} = g^{(\rm B)}_{\mu\nu} + h_{\mu\nu}\,.
\end{equation}
Under the Lorentz gauge condition $\nabla_{\mu}h^{\mu\nu}=0$ and for a transverse traceless tensor $g^{(\rm B)\mu\nu}h_{\mu\nu}=0$,
we have the propagation equation for GWs $h_{\mu\nu}$
\begin{equation}
\nabla^2h_{ij}+4(1-2\psi)\frac{\partial \psi}{\partial t}\frac{\partial h_{ij}}{\partial t}-(1-4\psi)\frac{\partial^2 h_{ij}}{\partial t^2}=0\,,
\end{equation}
where we have neglected higher order non-linear terms~\cite{Peters}. Using the eikonal approximation~\cite{Baraldo:1999ny}, the GW tensor can be represented as 
\begin{equation}
h_{ij} = u e_{ij}\,,
\end{equation}
where $e_{ij}$ is the polarization tensor of GWs and $u$ is a scalar field. Since the change of the polarization tensor by gravitational lensing is of the order of $\psi(t, {\bf{r}})\ll1$, we assume that the polarization tensor does not change during the propagation of GWs. Thus, we obtain a scalar wave equation as
\begin{equation}
\nabla^2u+4(1-2\psi)\frac{\partial \psi}{\partial t}\frac{\partial u}{\partial t}-(1-4\psi)\frac{\partial^2u}{\partial t^2}=0\,.
\end{equation}
We further recast the above equation into
\begin{equation}
c^2\nabla^2u+2b(c^2+1)\frac{\partial u}{\partial t}-\frac{\partial^2u}{\partial t^2}=0\,,\label{scalarwave}
\end{equation}
where $c$ is the speed of wave $c^2=1/(1-4\psi)$ with respect to a remote observer.
The parameter $b$ is defined as
$b=\frac{\partial \psi}{\partial t}$.

In this work, we consider the potential produced by a binary lens. Figure.~\ref{fig:twobody} shows the schematic of our system. We choose the origin of our coordinate system at the center of mass (barycenter). The trajectories of each object in the binary lens then can be described by  
\begin{equation}
\left\{
\begin{aligned}
{\bf{x}}_1(t)&=&\frac{r}{1+q}(\cos\omega t,\sin\omega t,0)\\
{\bf{x}}_2(t)&=&-\frac{qr}{1+q}(\cos\omega t,\sin\omega t,0)
\end{aligned}
\right.\,,
\end{equation}
where $r$ is the separation between two objects, $\omega=\sqrt{\frac{m_1+m_2}{r^3}}$ is the angular frequency of the orbit and $q=\frac{m_1}{m_2}\leq1$ is the mass ratio of the binary lenses\footnote{Here we adopt the convention that all the mass ratios are $\leq 1$.}. We take the potential generated by each object as 

\begin{equation}
\psi_i(t,\bf{x})=\left\{
\begin{aligned}
-\frac{m_i}{|{\bf{x}}-{\bf{x}}_i(t)|} ~~~~~~~~~~~~~~~~~|{\bf{x}}-{\bf{x}}_i(t)|>R_{s,i}\\
-m_i\frac{3R_{s,i}^2-|{\bf{x}}-{\bf{x}}_i(t)|^2}{2R_{s,i}^3} ~~~~~|{\bf{x}}-{\bf{x}}_i(t)|\leq R_{s,i}
\end{aligned} \right.\,,
\end{equation}	
where $R_{s,i}=2m_i$ is the Schwarzschild radius of object $i=1,2$ and $m_i$ is the mass of that object. Note that the waveform far away from the Schwarzschild radius does not depend on the form of potential well within $R_s$, which is shown explicitly in~\cite{He:2021hhl} by comparing numerical results of a static object with the analytical solution of a point source mass.
This is because GWs travel much faster in regions that are far away from the center than those close to it (the stronger potential, the smaller wave speed). As a result, for a distant observer, the lensed GWs mainly come from the outer regions while not from the inner regions. As such, regions near $R_s$ of the black hole, indeed, has a limited impact on the distant observer\footnote{For instance, in the relativistic case, the wave speed vanishes at the horizon of a black hole (infinite redshift surface), which means that no information can travel out of the horizon. As a result, the horizon of a black hole has a limited impact on a distant observer.}.   

In the weak field limit, the total potential is simply the superposition of the potential generated by each object in the binary lens 
\begin{align}
\psi=\psi_1+\psi_2 \nonumber\,.
\end{align}
Then its derivative is given by 
\begin{align}
\frac{\partial \psi}{\partial t}=\frac{\partial \psi_1}{\partial t}+\frac{\partial \psi_2}{\partial t} \,.\nonumber
\end{align}

\begin{figure*}[!htb]
\begin{center}
\includegraphics[width=2.1\columnwidth]{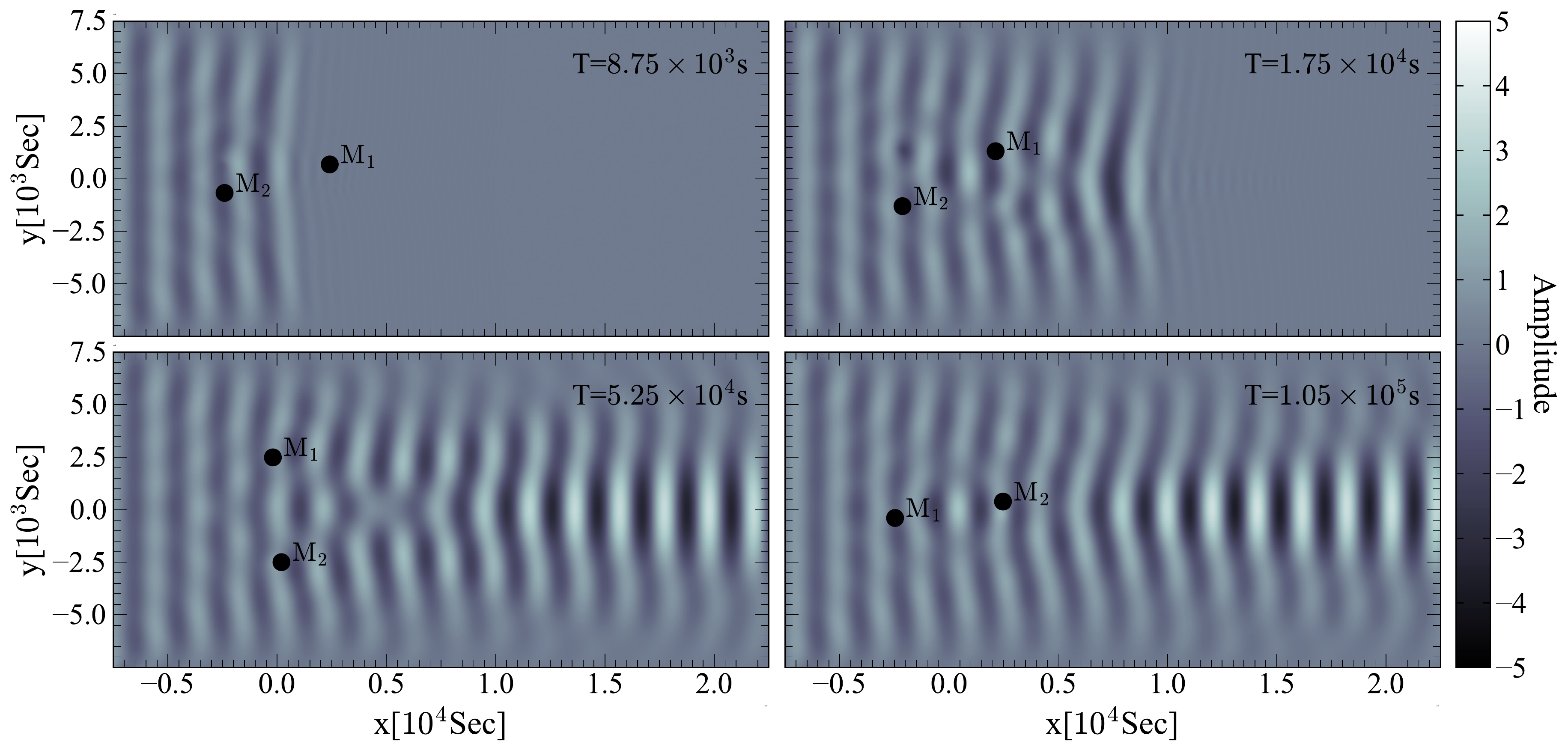}
\end{center}
\caption{The spatial waveform at different times for illustrative purposes. The snapshots are taken along the $x-y$ plane with $z=0$ [Sec]. The colour bar to the right shows the amplitude of GWs. The black dots indicate the positions of the binary black holes. The mass ratio of the binary lens is $1.0$ in this case. Unlike geometric optics, when GWs pass through the binary lens, there are no caustics but, instead, there is a strong beam along the optic axis ($x$-axis).}
\label{fig:amp_time}
\end{figure*}

In this work, we adopt the finite element method to solve Eq.~(\ref{scalarwave}). Our numerical simulations are based on the $\mathtt{GWsim}$~\cite{He:2021hhl} code, which is further based on the public available code $\mathtt{deal.ii}$~\cite{dealII90,dealII91,BangerthHartmannKanschat2007}. 
See Supplemental Material for the detailed numerical implements.

{\bf Numerical simulations} In this work, we assume that the GW source is generated by binary black holes with equal masses of $1\times 10^5 M_{\odot}$ at redshift $z=0.1$. The binary black holes are in their inspiral phase and rotate at a frequency of $5\times 10^{-4} {\rm Hz}$. There are about $27$ days for the binary black holes to coalescence. We assume that the length of our observing time is $T_{\rm obs}=2\times10^5$ [Sec], within which the binaries have an optimal signal-to-noise ratio (SNR) of $91.40$ with respect to the LISA sensitivity. Since within $T_{\rm obs}=2\times10^5$ [Sec], the change of the frequency of the binary system is less than $5\%$ \cite{Maggiore:2007ulw, Maggiore:2018sht}, we adopt a stationary phase approximation for the source binaries in our simulations. 

Then we assume that the source GWs are lensed by binary black holes with a total mass of $2.4\times 10^7 M_{\odot}$ at redshift $z=0.05$. There are about $5.6$ years for the lens black holes to coalescence. They also rotate at a stationary frequency of $5\times 10^{-6}$ Hz, which is outside of the sensitive range of the LISA detection band\cite{Baker:2019nia}. As such, the GWs generated by the lens itself can not be detected by LISA directly. Note that the parameters of our simulations are astrophysical motivated, as both the source and lens binaries are estimated to be well distributed with a population of $\sim 10^2-10^3$ within the redshift range of $0.01<z<0.1$ \cite{Katz:2019qlu}.

In this work, we choose the simulation domain as a cylinder with a radius of $7.5\times10^3$ [Sec] and a length of $3\times10^4$ [Sec]. The axis of the cylinder is taken along the $x$-axis ranging from  $-0.75\times10^4$ [Sec] to $2.25\times10^4$ [Sec] with the origin at the barycenter. The incident GWs travel normally along the $x$-axis. The simulation domain has a refinement of $2^7$ with a total of $2.1\times 10^7$ degrees of freedom (the sames as the total number of nodal points in the simulation domain). Given the tests presented in our previous work~\cite{He:2021hhl}, such resolution is sufficient for this work. In practice, because of the linearity of the wave equation Eq.~(\ref{scalarwave}) and the geometric unit, we simulate the scenarios with a re-scaling factor of $0.002$ for convenience. We run 5 simulations in total with different mass ratios as $q=\{1.0,0.5,0.2,0.125,0.1\}$. Each simulation uses $768$ CPU cores and takes about $73{\rm k}$ CPU hours.

\begin{figure}[]
\begin{center}
\includegraphics[width=0.5\textwidth]{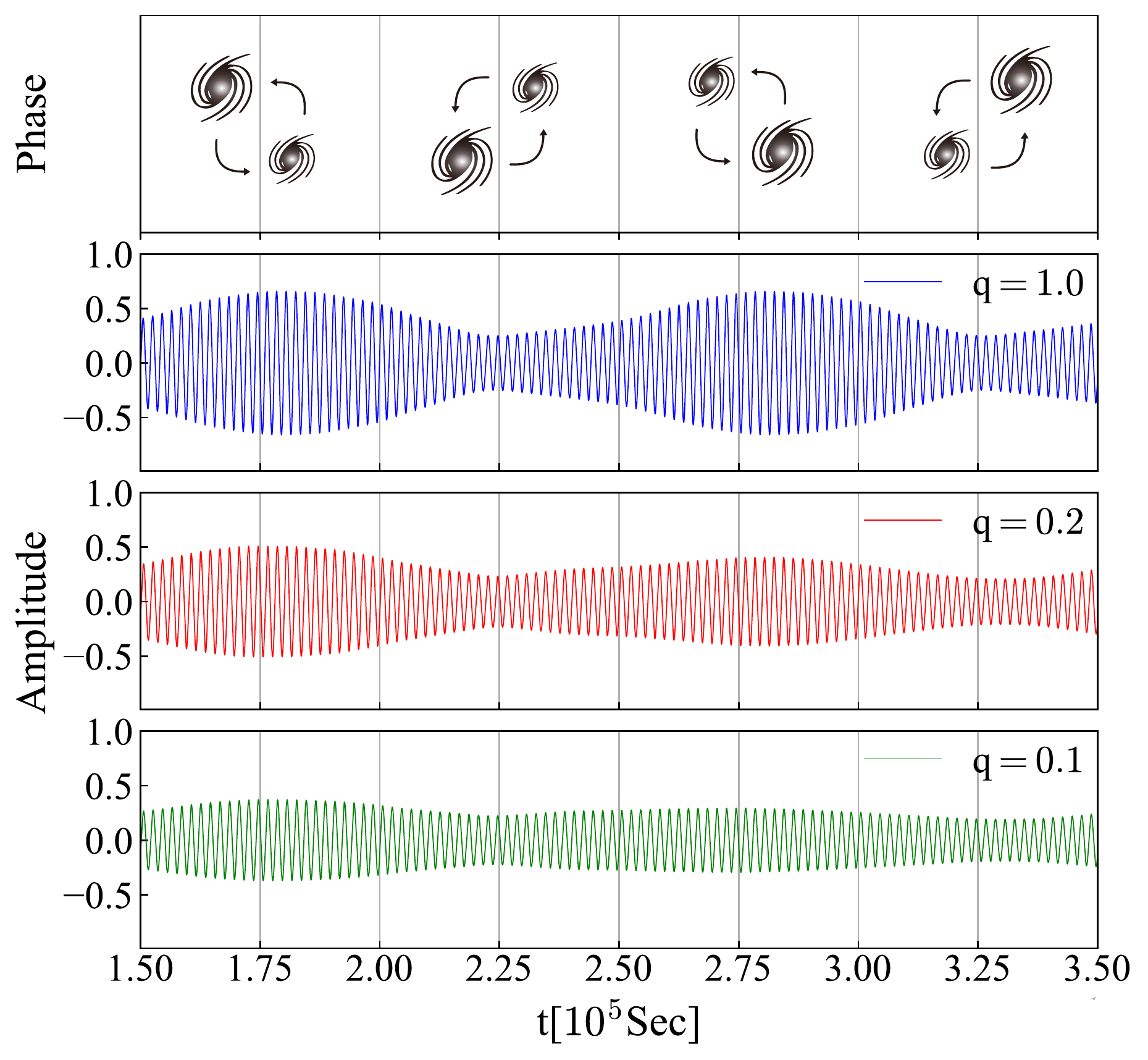}
\end{center}
\caption{The temporal waveform observed by an observer. The observer is located at $(x=+2.2\times10^4, y=+3.75\times10^3, z=0.00)$ [Sec]. The upper panel shows the different rotating phases of the binary lens. The lower three panels show the corresponding temporal waveforms for binary lens with mass ratios $\{1,0.2,0.1\}$. Unlike a single static lens, the amplitude of the source GWs are periodically modulated by the binary lens.}
\label{fig:illu+time}
\end{figure}

Figure.~\ref{fig:amp_time} shows the spatial waveforms at different times for illustrative purposes. The snapshots are taken along the $x-y$ plane with $z=0$ [Sec]. The colour bar to the right shows the amplitude of waves. The black dots indicate the positions of the binary black holes. Unlike geometrical optics, when GWs pass through the binary lens, they do not form caustics but, instead, they form a strong beam of signals along the optic axis ($x$-axis).

Figure.~\ref{fig:illu+time} shows the temporal waveform observed by an observer at $(x=+2.2\times10^4, y=+3.75\times10^3, z=0.00)$ [Sec]. The total evolution time of our simulation is $3.5\times10^5$ [Sec], which is about $\approx 1.75$ cycle of the binary lens. The upper panel of Figure.~\ref{fig:illu+time} shows the different rotation phases of the binary lens. The corresponding temporal waveforms at the observer are shown in the lower three panels for different mass ratios $q=\{1,0.2,0.1\}$, respectively. The time is from $1.5\times10^5$[Sec] to $3.5\times10^5$ [Sec] , which covers one entire cycle $T_{\rm obs}=2\times10^5$ [Sec] of the binary lens. Unlike in the case of a single static lens, the most prominent feature of the binary lens is that the amplitude of GWs from the source are periodically modulated by the binary lens.

\begin{figure}[]
\begin{center}
\includegraphics[width=0.5\textwidth]{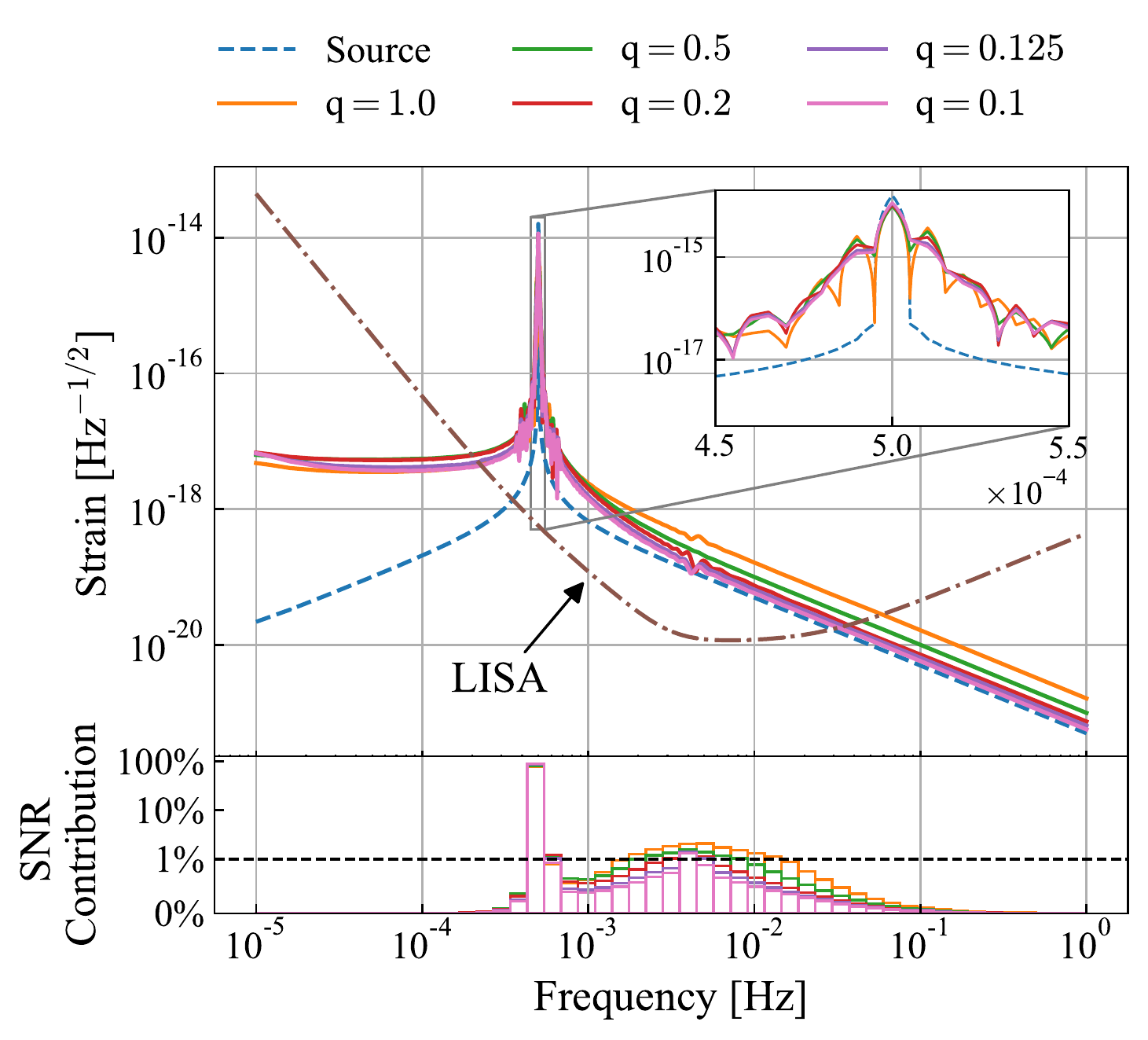}
\end{center}
\caption{Upper panel: The power spectrum density of the lensed GW signals for binary lens with different mass ratios $\{1, 0.5, 0.2,0.125,0.1\}$ (solid lines). Here, the amplitudes of GW signals are normalized as the cases without lensing magnification $\mu=1$. The light blue dashed line is for the unlensed GW signals. The brown dash-dotted line represents the sensitivity curve of LISA. The inset shows the near-peak region, which highlights the $4.9\times10^{-4}$Hz and $5.1\times10^{-4}$Hz sidebands that correspond to the convolution of the lens and source frequencies. Lower panel: The relative contribution to SNR with respect to different bins of frequency. The horizontal dashed line indicates the $1\%$ threshold, below which the linear scaling is applied. SNR mainly comes from the near-peak region.  }
\label{fig:FFT}
\end{figure}

{\bf SNR with respect to LISA}
We first calculate the power spectral density (PSD) of the 1-D time-domain GW signals. To do this, we use the Fast Fourier Transform (FFT) technique to perform the Fourier Transform of the time-domain waveform. The integration is over the time-limited GW signals $T_{\rm obs}=2\times10^5$ [Sec]. As a result, the frequency resolution of the power spectral is $1/T=5\times10^{-6} $ Hz.

Figure.~\ref{fig:FFT} shows the PSD of the lensed GW signals. Compared with the unlensed ones that are simply the sinusoid's spectrum (blue dashed line), in the near-peak zone of the lensed signals (solid lines), the nearest two peaks on both sides of the main peak are merged into broader sidebands (shown in the inset). These sidebands locate exactly at the $4.9\times10^{-4}$ Hz and $5.1\times10^{-4}$ Hz, which correspond to the convolution of the lens and source frequencies $f_{\rm sideband}=f_{\rm source}\pm 2f_{\rm lens}$. The lower panel of Figure.~\ref{fig:FFT} shows the relative contribution to SNR with respect to different bins of frequency. SNR mainly comes from the near-peak region.

To estimate the detectability of this scenario, we calculate the matched filter SNR with respect to the LISA mission-required 2-arm sensitivity using ~\cite{Haris:2018vmn, SciRD}
\begin{align}
\left(s|h\right)\equiv2\sum_{D} \int_{0}^{\infty} \frac{s(f)^*h(f)+s(f)h(f)^*}{S_{D}(f)} d f\,.
\end{align}
The sum in above equation is over all detectors.
The optimal detection SNR $\rho_{{\rm opt}}$ is obtained when the template $h$ matches the signal $s$. For the lensed signals, $\rho_{{\rm opt}}$ is given by
\begin{equation}
\rho_{{\rm opt}}^{\text {lens }}=\sqrt{\left(h^{\rm lens}|h^{\rm lens}\right)}=\left(2\sum_{D} \int_{f_{\text {low }}}^{\infty} \frac{h_{D}^{\text {lens }}(f)^{2}}{S_{D}(f)} d f\right)^{1 / 2}\,,\label{SNR_LISA}
\end{equation}
where $h_{D}^{\text {lens }}(f):=F_{+, D}(\alpha, \delta, \psi) h_{+}^{\text {lens }}(f)+F_{\times, D}(\alpha, \delta, \psi)$ $h_{\times}^{\text {lens }}(f)$ denotes the observed signal in detector $D$. $F_{+, D}$ and $F_{\times, D}$ are the antenna's pattern functions, which depend on the source position $\alpha, \delta$ and polarization angle $\psi$. Here we adopt the all-sky averaged inclination factor $5/4$ for the pattern functions~\cite{Finn:1995ah}. Moreover, we use the latest sensitivity curve $S_{D}(f)$ of LISA provided by ~\cite{Gair:2022knq, Thrane:2013oya} to calculate the $\rho_{{\rm opt}}^{\text {lens }}$. 

\begin{figure}[]
\begin{center}
\includegraphics[width=0.5\textwidth]{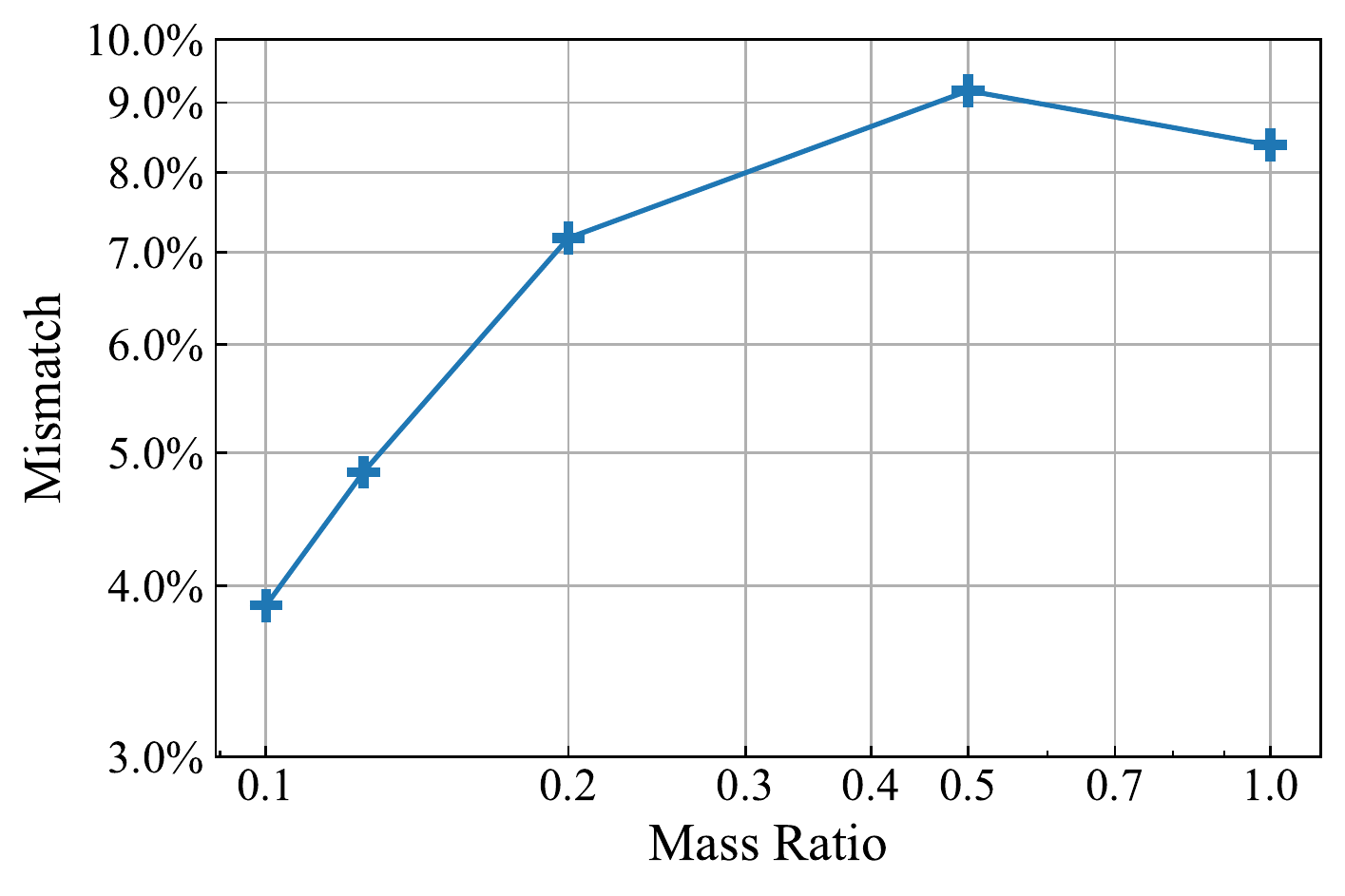}
\end{center}
\caption{The {\it mismatch} $1-\mathcal{M}$ between the lensed and unlensed templates with respect to different mass ratios. {\it Mismatches} are greater than $5\%$ for mass ratios $q>0.125$. Since {\it mismatch} does not depend on the amplitude of waveforms, the differences between the lensed and unlensed waveforms are substantial. }
\label{fig:mismatch}
\end{figure}

To highlight the relative SNR differences between the lensed $h^{\rm lens}$ and unlensed waveforms $h^{\rm T}$, we measure the 
{\it mismatch} ($1-\mathcal{M}$) between $h^{\rm lens}$ and $h^{\rm T}$, where $\mathcal{M}$, namely {\it match}, is defined by
\begin{align}
\mathcal{M}[h^{\rm lens},h^{\rm T}]\equiv\frac{\left(h^{\rm lens}|h^{\rm T}\right)}{\sqrt{\left(h^{\rm lens}|h^{\rm lens}\right)\left(h^{\rm T}|h^{\rm T}\right)}}\,.
\end{align}
Note that the {\it mismatch} defined in the above way does not depend on the amplitude of waveforms, as amplitudes are cancelled out in the above expression. As such, the {\it mismatch} depends only on the shape of the waveform.

\begin{table}[!t]  \caption{SNR for the source and lensed signals.}
 \label{table:SNR}
 \begin{tabular}{cccccccc}  
\toprule   
  Mass ratio & 1 & 0.5 & 0.2 & 0.125 & 0.1 & Source\\  
\midrule   
   SNR & 71.39 & 67.86 & 72.21 & 75.00 & 76.37 & 95.26\\  
  \bottomrule  
\end{tabular}
\end{table}

Table~\ref{table:SNR} shows the SNR of the lensed and unlensed GWs for different mass ratios of the binary lens. The SNR of the source is $91.40$, which is much larger than the detection threshold of LISA ${\rm SNR}\ge 8$. Figure.~\ref{fig:mismatch} shows the {\it mismatch}  ($1-\mathcal{M}$) between the lensed and unlensed waveforms with respect to different mass ratios. The {\it mismatch} is greater than $5\%$ for mass ratios $q>0.125$. 

{\bf Summary and discussions}
In this paper, we have studied the time-domain waveforms of GWs passing through a lensing system that consists of binary black holes as the lens using numerical simulations. Unlike the single static lens, the lensing potential of the binary lens evolves with time, which can modulate periodically the amplitude of GW signals that pass through it. This phenomenon is similar to AM in radio communication.

In the frequency-domain, a prominent feature of the amplitude modulated GWs is that, in the near-peak region, the nearest two peaks on both side of the main peak are merged into broader sidebands, which locate at the positions that correspond to the convolution of the lens and source frequencies $f_{\rm sideband}=f_{\rm source}\pm 2f_{\rm lens}$. As a result, even the frequency of the binary lens itself is too low to be detected by LISA, the sidebands due to AM can still be within the sensitive range of the LISA detection band. Moreover, since SNR mainly comes from the near-peak region, such sidebands can lead to significant {\it mismatch} of SNR between the lensed and unlensed GWs.

Using numerical simulations with astrophysical motivated parameters, we have calculated the SNR of the waveforms after being lensed by a binary lens. We find that the {\it mismatch} between the lensed and unlensed GWs are substantial, which are more than $5\%$ for mass ratios $q>0.125$. This is in contrast with the case of a single static lens. Our results thus demonstrate that the standard GR templates can not be directly used to search the event of a binary lens. Templates that take into account the effect of binary lens have to be used. The substantial differences between the binary lensed and unlensed waveforms, however, in turn, provide a robust way to identify the lensing event for the LISA project in the future. 

Our findings also demonstrate the possibilities of using AM in the LISA project to detect binary lenses with frequencies beyond its detection band. Using the techniques of demodulation in radio communication, it is also possible to extract the mass ratio of the binary lens from the sidebands of the lensed power spectrum. This may provide a potential way to infer the individual mass of the binary lens. The detailed discussions on this topic will be presented in our future work.

\vspace{5mm}
\noindent {\bf Acknowledgments} 
We thank Chengjiang Yin and Xiangyu Xu for helpful discussions.
The numerical calculations in this paper have been done on the computing facilities in the High Performance Computing Center (HPCC) of Nanjing University. This work is supported by the National Key R$\&$D Program of China (Grant No. 2021YFC2203002, No. 2021YFC2203003), the National Natural Science Foundation of China (Grants No. 12075116, No. 12150011, No. 12005084, No. 12047501), the science research grants from the China Manned Space Project (Grant NO.CMS-CSST-2021-A03).

\bibliography{binarylensing_v2}% Produces the bibliography via BibTeX.

\newpage

\section*{Supplementary Materials}

The finite element method is based on the weak formulation of the wave equation
\begin{eqnarray}
\langle\phi,\frac{\partial u}{\partial t}\rangle_{\Omega} &\equiv&\langle\phi,v\rangle_{\Omega}\,,\\
\nonumber
\langle\phi,\frac{\partial v}{\partial t}\rangle_{\Omega}&=&-\langle\nabla(c^2\phi),\nabla u\langle_{\Omega}-\langle c\phi,\frac{\partial u}{\partial t}\rangle_{\partial\Omega}\\
&&+\langle 2b(c^2+1)\phi,\frac{\partial u}{\partial t}\rangle_{\Omega}\,,
\end{eqnarray}
where $\phi$ is a test function and we use the notion 
\begin{align}
\langle f,g \rangle_{\Omega}=\int_{\Omega}f(x)g(x)dx\,,\nonumber
\end{align}
for convenience. In the second equality, we have imposed an absorbing boundary condition 
\begin{align}
\hat{n}\cdot\nabla u=-\frac{1}{c}\frac{\partial u}{\partial t}\,
\end{align}
on the surfaces of our simulation domain. However, the absorbing boundary condition does not apply to the domain surface where GWs enter our simulation domain. 

In this work, we discretize the time variable first, following the Rothe's method
\begin{eqnarray}
\langle \phi,\frac{u^n-u^{n-1}}{k}\rangle_{\Omega}&=&\langle \phi,\theta v^n+(1-\theta)v^{n-1}\rangle_{\Omega}\,,\label{equ}\\
\nonumber
\langle \phi,\frac{v^n-v^{n-1}}{k}\rangle_{\Omega}&=&-\langle\nabla(c^2\phi),\nabla[\theta u^n+(1-\theta)u^{n-1}]\rangle_{\Omega}\\
\nonumber
&&+\langle 2b(c^2+1)\phi,\theta v^n+(1-\theta)v^{n-1}\rangle_{\Omega}\\
&&-\langle c\phi,\frac{u^n-u^{n-1}}{k}\rangle_{\partial\Omega}\,,\label{eqv}
\end{eqnarray}  
where the superscript $n$ indicates the number of a time step and $k=t_n-t_{n-1}$ is the length of the present time step. In this work, we choose $\theta = \frac{1}{2}$, which is called the Crank-Nicolson scheme. This scheme is implicit. An advantage of the implicit scheme is that it is numerically stable. In the above equations, $b$ and $c$ are functions of time. Since their values are known at every time, they can be treated as known parameters. 

Next, we discretize the spatial variables using the finite element method. At each time step, we expand $u^n$, $v^n$, $u^{n-1}$ and $v^{n-1}$ in terms of the shape function $\phi_i$ on each element
\begin{equation}
\left\{
\begin{aligned}
u^n&\approx\sum_iU^n_i\phi_i\\
v^n&\approx\sum_iV^n_i\phi_i\\
u^{n-1}&\approx\sum_iU^{n-1}_i\phi_i\\
v^{n-1}&\approx\sum_iV^{n-1}_i\phi_i
\end{aligned}\right.\,,
\end{equation}
where $U^n_i\,,V^n_i\,,U^{n-1}_i\,,V^{n-1}_i$ are unknown constant coefficients. Inserting the above expressions back into Eqs.~(\ref{equ},\ref{eqv}), the unknown coefficients form a group of linear systems
\begin{align}
&\left[M+k^2\theta^2(A+D)+k\theta(B-C)\right]U^n\nonumber\\
=& \left[M-k^2\theta (1-\theta)(A+D)+k\theta (B-C)\right]U^{n-1}\nonumber\\
&+kMV^{n-1}\,,\label{eqUM}\\
&\left[M+k^2\theta^2(A+D)+k\theta(B-C)\right]V^n\nonumber\\
=&\left[M-k^2\theta(1-\theta)(A+D)-k(1-\theta)(B-C)\right]V^{n-1}\nonumber\\
&-k(A+D)U^{n-1}\,,\label{eqVM}
\end{align}
where the elements of the matrices are defined by
\begin{equation}
\left\{
\begin{aligned}
A_{ij}&=\langle c^2\nabla\phi_i,\nabla\phi_j\rangle_{\Omega}\\
B_{ij}&=\langle c\phi_i,\phi_j\rangle_{\partial\Omega}\\
C_{ij}&=\langle 2b(c^2+1)\phi_i,\phi_j\rangle_{\Omega}\\
D_{ij}&=\langle \nabla(c^2)\phi_i,\nabla\phi_j\rangle_{\Omega}\\
M_{ij}&=\langle \phi_i,\phi_j\rangle_{\Omega}
\end{aligned}\right.\,.
\end{equation}
$U^n$ and its time derivative $V^n$ in Eqs.~(\ref{eqUM},\ref{eqVM}) at a time step $t_n$ are independent to each other. They only depend on $U^{n-1}$ and $V^{n-1}$ at a previous time step $t_{n-1}$.

Equations~(\ref{eqUM},\ref{eqVM}) can be solved using an iterative method. Since the matrices are not symmetric, we adopt the GMRES (a generalized minimal residual algorithm for solving non-symmetric linear systems) method, which does not require any specific properties of the matrices. 

\end{document}